\documentclass[letterpaper,twocolumn,fleqn,superscriptaddress,preprintnumbers,prd]{revtex4}

\usepackage{graphicx,color}
\usepackage{amsmath,amssymb,amsfonts}
\usepackage{stackrel}
\usepackage{tabularx}
\newcolumntype{Y}{>{\centering\arraybackslash}X}
\usepackage{booktabs}
\AtBeginDocument{
    \heavyrulewidth=.08em
        \lightrulewidth=.05em
        \cmidrulewidth=.03em
        \belowrulesep=.65ex
        \belowbottomsep=0pt
        \aboverulesep=.4ex
        \abovetopsep=0pt
        \cmidrulesep=\doublerulesep
        \cmidrulekern=.5em
        \defaultaddspace=.5em
}
\usepackage{xcolor}
\usepackage{hyperref}
\hypersetup{
    colorlinks=true,
        linkcolor={blue!80!black},
        citecolor={blue!80!black},
        urlcolor={blue!80!black} 
}

\providecommand{\openone}{\leavevmode\hbox{\large1\kern-7.3pt\normalsize1}}
\newcommand{\be}{\begin{equation}}
\newcommand{\ee}{\end{equation}}
\newcommand{\ba}{\begin{eqnarray}}
\newcommand{\ea}{\end{eqnarray}}

\begin{document}

\title{Gravitational-wave constraints on the neutron-star-matter Equation of State}

\preprint{CERN-TH-2017-236, HIP-2017-30/TH}
\author{Eemeli Annala}
\affiliation{Department of Physics and Helsinki Institute of Physics, P.O.~Box 64, FI-00014 University of Helsinki, Finland}
\author{Tyler Gorda}
\affiliation{Department of Physics and Helsinki Institute of Physics, P.O.~Box 64, FI-00014 University of Helsinki, Finland}
\author{Aleksi Kurkela}
\affiliation{Theoretical Physics Department, CERN, Geneva, Switzerland and \\ Faculty of Science and Technology, University of Stavanger, 4036 Stavanger, Norway}
\author{Aleksi Vuorinen}
\affiliation{Department of Physics and Helsinki Institute of Physics, P.O.~Box 64, FI-00014 University of Helsinki, Finland}

\begin{abstract}

\noindent The LIGO/Virgo detection of gravitational waves originating from a neutron-star merger, GW170817, has recently provided new stringent limits on the tidal deformabilities of the stars involved in the collision. Combining this measurement with the existence of two-solar-mass stars, we generate a generic family of neutron-star-matter Equations of State (EoSs) that interpolate between state-of-the-art theoretical results at low and high baryon density. Comparing the results to ones obtained without the tidal-deformability constraint, we witness a dramatic reduction in the family of allowed EoSs. Based on our analysis, we conclude that the maximal radius of a 1.4-solar-mass neutron star is 13.6 km, and that smallest allowed tidal deformability of a similar-mass star is $\Lambda(1.4 M_\odot) = 120$.

\end{abstract}

\maketitle

%%%%%%%%%%%%%%%%%%%%%%%%%%%%%%%%%%%%%%%%%%%
\section{Introduction}
%%%%%%%%%%%%%%%%%%%%%%%%%%%%%%%%%%%%%%%%%%%

The collective properties of the strongly-interacting dense matter found inside neutron stars (NS) are notoriously difficult to predict \cite{Lattimer:2004pg,Brambilla:2014jmp}. While the Sign Problem prevents lattice Monte-Carlo simulations at nonzero chemical potentials \cite{deForcrand:2010ys}, nuclear-theory tools such as Chiral Effective Theory (CET) are limited to sub-saturation densities \cite{Tews:2012fj} and perturbative QCD (pQCD) becomes reliable only at much higher densities \cite{Kurkela:2009gj}. No controlled, first-principles calculations are applicable at densities encountered inside the stellar cores. 

Despite these difficulties, it is possible to obtain robust information on the properties of NS matter at core densities. In particular, the requirement that the Equation of State (EoS) must reach its known low- and high-density limits while behaving in a thermodynamically consistent fashion in between poses a strong constraint on its form. This was demonstrated, e.g.,~in \cite{Kurkela:2014vha,Fraga:2015xha}, where a family of EoSs was constructed that interpolate between a CET EoS below saturation density and a pQCD result at high densities. This family quantifies the purely theoretical uncertainty on the EoS at intermediate densities, but the quantity can be further constrained using observational information about the macroscopic properties of NSs. 

The first significant constraint for the EoS comes from the observation of two-solar-mass (2$M_{\odot}$) stars \cite{Demorest:2010bx,Antoniadis:2013pzd}, implying that the corresponding mass-radius curve must support massive enough stars, $M_{\rm max} > 2 M_\odot$. This requires that the EoS be stiff enough, which in combination with the fact that the high-density EoS is rather soft (with $c_s^2 \lesssim 1/3$; $c_{s}$ is the speed of sound) limits the possible behavior of the quantity at intermediate densities. In particular, it was shown in \cite{Kurkela:2014vha,Fraga:2015xha} that---upon imposing the $2M_{\odot}$ constraint---the current uncertainty in the EoS when expressed in the form $p(\mu_{\text{B}})$, with $p$ being the pressure and $\mu_{\text{B}}$ the baryon chemical potential, is $\pm 40\%$ at worst. 

%%%%%%%%%%%%%%%%%%%%%%%%%%%%%%%%%%%%%%%%%%%%%%%%%%%%%%%%%%%%
\begin{figure}
\includegraphics[width=0.48\textwidth]{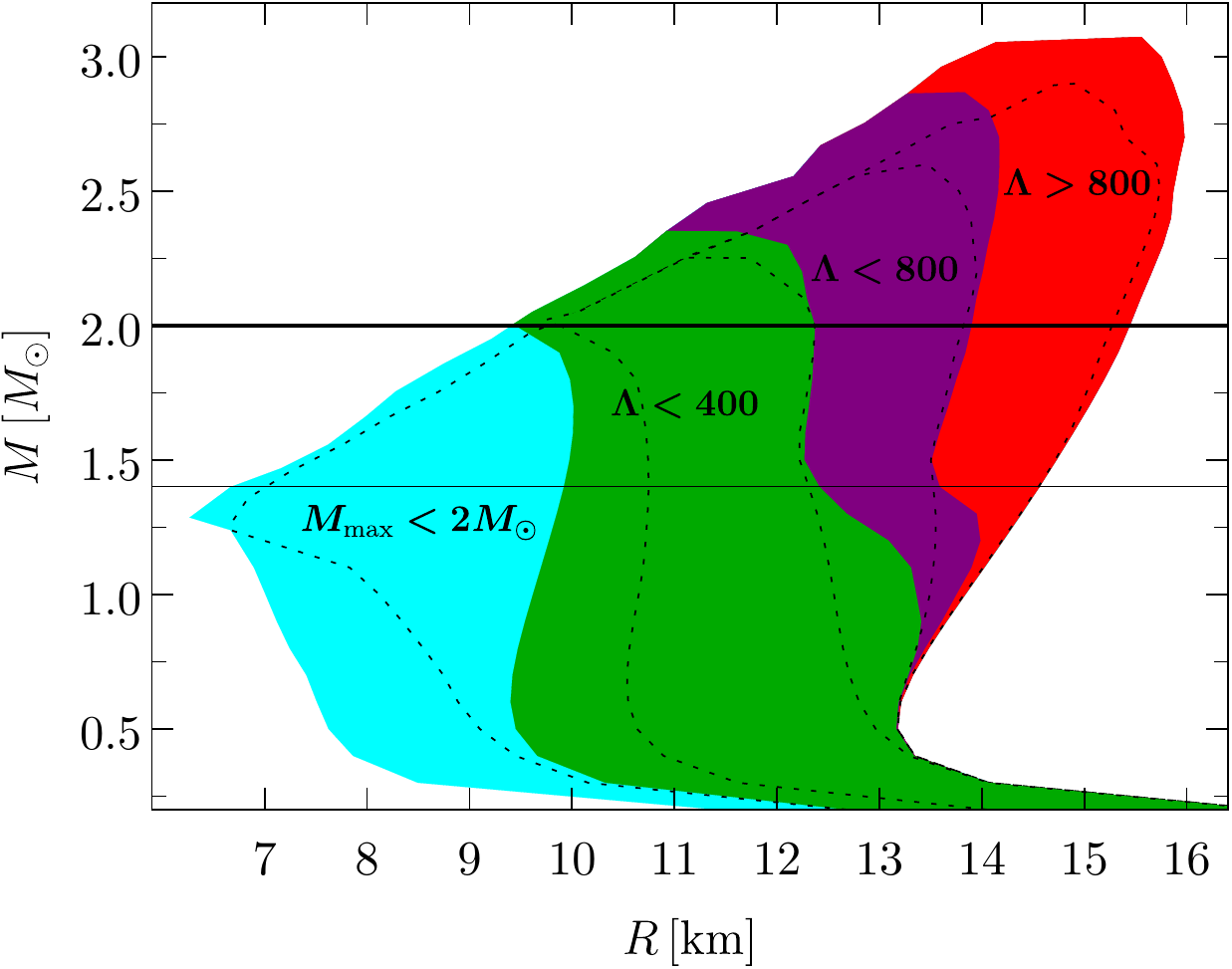}
\caption{
The mass-radius clouds corresponding to our EoSs. The cyan area corresponds to EoSs that cannot support a $2M_\odot$ star, while the rest denote EoSs that fulfill this requirement and in addition have $\Lambda(1.4M_\odot) < 400$ (green), $400 < \Lambda(1.4M_\odot) < 800$ (violet), or $\Lambda(1.4M_\odot) > 800$ (red), so that the red region is excluded by the LIGO/Virgo measurement at 90\% credence. This color coding is used in all of our figures. The dotted black lines denote the result that would have been obtained with tritropic interpolation only.}
\label{MRplot}
\end{figure}
%%%%%%%%%%%%%%%%%%%%%%%%%%%%%%%%%%%%%%%%%%%%%%%%%%%%%%%%%%%%

On 16 October 2017, the LIGO and Virgo collaborations reported the first event, GW170817, where a gravitational-wave (GW) signal was observed from a merger of two compact stars \cite{TheLIGOScientific:2017qsa}. Remarkably, this first set of GW data already offers a second constraint for the behavior of NS matter. The inspiral phase of a NS-NS merger creates strong tidal gravitational fields that deform the multipolar structure of the stars, which in turn leaves a detectable imprint on the observed gravitational waveform of the merger. This effect can be quantified in terms of the so-called tidal deformabilities  $\Lambda_{i} = (2 / 3) k^{(i)}_{2} [ (c^{2} / G) R_{i} / M_{i}]^{5}$ of the stars, where $k^{(i)}_{2}$ is the second Love number, $R_{i}$ the radius, and $M_{i}$ the mass of the $i$th star \cite{Flanagan:2007ix,Hinderer:2007mb}. Assuming a low-spin prior for both stars involved in the merger (for details, see \cite{TheLIGOScientific:2017qsa}), LIGO and Virgo quote that $\Lambda(1.4 M_\odot ) < 800$ with a credence level of 90\%. Since $\Lambda$ is a quantity closely related to the EoS of stellar matter, this measurement provides another constraint for NS matter.

In this paper, we revisit the problem of generating the most generic family of NS-matter EoSs consistent with all robust theoretical and observational constraints. It is seen that the inclusion of the new upper bound on $\Lambda$ significantly constrains the EoS and quantities derived from it, such as the mass-radius relation. As hard EoSs lead to stars with large radii and large tidal deformabilities, an upper bound on $\Lambda$ brackets the EoS from a direction opposite the 2$M_{\odot}$ observation. These effects are summarized in Fig.~\ref{MRplot}, which shows that while the $2M_\odot$ constraint implies $R(1.4M_\odot) > 9.9$~km, the new limit from the $\Lambda$ measurement reads $R(1.4M_\odot) < 13.6$~km.

It should be noted that similar---and in some cases more stringent---limits on NS properties have been reported elsewhere. These other studies, however, are either based on a small set of individual EoSs (see e.g.~\cite{Bauswein:2017vtn,Radice:2017lry,Yamamoto:2017wre}), interpret observational data in a way that contains modeling uncertainties \cite{Guillot:2013wu,Bogdanov:2016nle,Steiner:2017vmg,Nattila:2015jra,Nattila:2017wtj}, or apply Bayesian inference to assess the credence of different EoSs based on a theoretical prior \cite{Guillot:2013wu,Bogdanov:2016nle,Nattila:2015jra,Nattila:2017wtj}. Some exceptions to this are the works of Hebeler et al.,~which extrapolates a CET EoS to higher densities \cite{Hebeler:2013nza}, and Kurkela et al.~\cite{Kurkela:2014vha} and Gorda \cite{Gorda:2016uag}, which additionally include a pQCD constraint at high density. In comparison to these studies, our current work implements a more generic interpolation of the EoS and also implements the recent LIGO/Virgo limit on $\Lambda$.

%%%%%%%%%%%%%%%%%%%%%%%%%%%%%%%%%%%%%%%%%%%
\section{Setup}
%%%%%%%%%%%%%%%%%%%%%%%%%%%%%%%%%%%%%%%%%%%

As discussed, e.g.,~in \cite{Hebeler:2013nza}, well-established nuclear-physics methods are sufficient to reproduce the EoS of cold, electrically-neutral, strongly-interacting matter in beta equilibrium---NS matter for short---up to approx.~the nuclear saturation density of $n_s\approx 0.16$ baryons per fm$^{3}$. Around this value, however, the underlying uncertainties in most modern calculations start to rapidly increase, so that the estimated theoretical error in, e.g.,~the state-of-the-art CET EoS of \cite{Tews:2012fj,Hebeler:2013nza} becomes $\pm 24\%$ at a density of $n=1.1 n_s$. In our calculation, we choose as the EoS below this density either the ``hard'' or ``soft'' EoS of \cite{Hebeler:2013nza}, which correspond to the most extreme EoSs allowed at low densities. 

For the EoS of deconfined quark matter at high density, we employ the NNLO pQCD result of \cite{Kurkela:2009gj}, which becomes increasingly accurate with larger density due to the asymptotic freedom of QCD. Here, the uncertainty level of $\pm 24\%$ is reached at $\mu_\text{B}=2.6$~GeV, corresponding to densities of approx.~40$n_s$. This result is parameterized by the renormalization scale parameter $X\in [1,4]$, introduced in \cite{Fraga:2013qra}, whose variation generates the uncertainty band. 

Between the regions of validity of CET and pQCD, one should allow the EoS to behave in any thermodynamically consistent manner. Following and extending the approach of \cite{Kurkela:2014vha} (cf.~also \cite{Hebeler:2013nza}), we form our EoSs by dividing the density interval from $n=1.1n_s$ to $\mu_\text{B}=2.6$~GeV into segments in $\mu_\text{B}$ and by assuming that within each $\mu_{i} < \mu_\text{B} < \mu_{i+1}$ the EoS has a polytropic form $p_{i}(n)=\kappa_i n^{\gamma_i}$. These segments are connected to each other by assuming that both the pressure and energy density behave continuously at each matching point. For $N$ segments, we have $N-1$ independent matching chemical potentials $\mu_i$ and $N$ independent polytropic indices $\gamma_i$, two of which are determined by matching to the low- and high-density EoSs, leaving $2N -3$ free parameters for given low- and high-density EoSs. To confirm that our results are independent of the interpolation, we consider polytropes which consist of either three (tritropes) or four (quadrutropes) polytropic segments, later verifying that the corresponding results agree at a sufficient accuracy. 

To obtain our ensemble of EoSs, we pick random values for the remaining free parameters from uniform distributions $\gamma_i \in [0 , 15]$, $\mu_i \in [\mu_{B}(1.1n_{s}) , 2.6\,{\rm GeV}]$, and $X \in [1 , 4]$, and choose the same number of ``soft'' or ``hard'' low-density EoSs. Note that by not enforcing any nontrivial lower limit on the $\gamma_i$, we effectively allow for a first order phase transition at any of the matching points. These random values sometimes results in instances where either no smooth solution is found or the resulting EoS is superluminal, $c_s^2>1$. We drop such solutions. 
We furthermore improve the coverage of parameter values by iteratively sampling parameters close to the values of extremal EoSs that define the boundaries of our allowed regions. This process leaves us finally with ensembles of 90,000 tri- and 170,000 quadrutropic EoSs.

Having constructed the family of EoSs, we next enforce the $2 M_{\odot}$ and $\Lambda$ constraints. This is done by simulateneously solving the mass-radius relations and tidal deformabilities for non-rotating stars, following a setup explained in some detail in Ref.~\cite{Hinderer:2009ca}.

%%%%%%%%%%%%%%%%%%%%%%%%%%%%%%%%%%%%%%%%%%%
\section{Results}
%%%%%%%%%%%%%%%%%%%%%%%%%%%%%%%%%%%%%%%%%%%

We proceed now to present and analyze the obtained EoS families. The allowed ranges of EoS parameters and the resulting macroscopic NS properties are summarized in Table I below. Unless stated otherwise, all of the figures shown are prepared with the full set of tri- and quadrutropic EoSs.

\subsection{Constraints on astrophysical observables}

\begin{figure}
\includegraphics[width=0.48\textwidth]{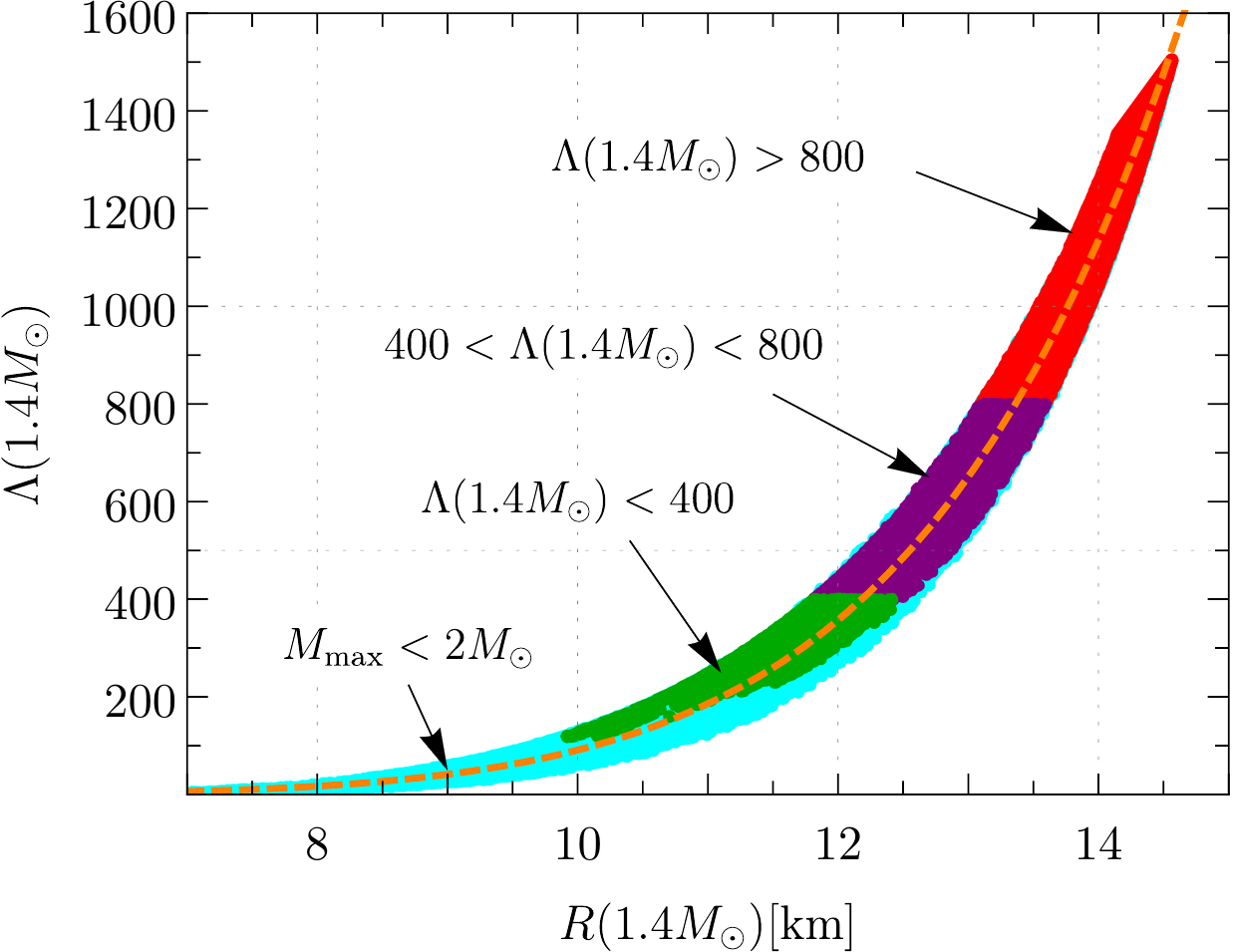}
\caption{The $\Lambda$ values for stars with $M=1.4 M_\odot$ as functions of the corresponding radius. The color coding follows Fig.~\ref{MRplot}, while the orange line $\Lambda = 2.88 \times 10^{-6} (R/\text{km})^{7.5}$ has been included just to guide the eye. 
}
\label{Lambda_vs_R}
\end{figure}

Since $\Lambda$ measures the deviation of the stellar gravitational field from that of a point-like mass, it is natural to expect that larger-radius stars possess larger $\Lambda$. In Fig.~\ref{Lambda_vs_R}, we indeed see a tight correlation between $R$ and $\Lambda$ for our ensemble of EoSs, each determined for stars with $M=1.4 M_\odot$. To a rather good accuracy, all tidal deformabilities are observed to follow the empirical function $\Lambda(R) = 2.88\times 10^{-6} (R/{\text{km}})^{7.5}$, shown as the orange dashed line in this figure. 

Due to the correlation between $R$ and $\Lambda$, the LIGO/Virgo measurement leads to a strong constraint on the possible radii of NSs: the 90\% limit of $\Lambda(1.4M_\odot) < 800$ \cite{TheLIGOScientific:2017qsa} directly translates into an upper limit of $R(1.4 M_\odot) < 13.6\; {\rm km}$. Should this bound be tightened to $\Lambda(1.4M_\odot) < 400$ in the future (as roughly suggested by the 50\% contours of Fig.~5 of \cite{TheLIGOScientific:2017qsa}), the constraint would further tighten to $R(1.4 M_\odot) < 12.5\, {\rm km}$. 

We note that this maximal-radius constraint is unaffected by the proposed limit of $M_{\max} < 2.16 M_{\odot}$ \cite{Margalit:2017dij,Rezzolla:2017aly,Ruiz:2017due}, stemming from the resulting kilonova observations associated with the GW170817 event \cite{TheLIGOScientific:2017qsa,Coulter:2017wya,GBM:2017lvd,Monitor:2017mdv,Cowperthwaite:2017dyu,Soares-Santos:2017lru,Nicholl:2017ahq,Margutti:2017cjl,Chornock:2017sdf}. This is because, for a given radius, there are many EoSs whose maximal mass is smaller than that of the EoS with the highest maximal mass.

%%%%%%%%%%%%%%%%%%%%%%%%%%%%%%%%%%%%%%%%%%%%%%%%%%%%%%%%%%%%
%%%%%%%%%%%%%%%%%%%%%%%%%%%%%%%%%%%%%%%%%%%%%%%%%%%%%%%%%%%%

While the LIGO/Virgo limit on $\Lambda$ favors soft EoSs, the 2$M_{\odot}$ constraint favors hard EoSs, thus setting a restrictive bound for the quantity.  For those EoSs that do support a 2$M_{\odot}$ star, the tidal deformabilities are found to take values in the range $\Lambda(1.4M_\odot) \in [ 120, 1504 ]$, implying that values smaller than 120 can be firmly ruled out. A further investigation shows that the minimal allowed values of $\Lambda$ depend strongly on the low-density EoS: those interpolated EoSs that are built with a soft hadronic component correspond to $\Lambda \in [120,1353]$, while those with a hard low-density part correspond to $\Lambda \in [161, 1504]$. Similarly, the 2$M_{\odot}$ constraint is seen to lead to a stringent limit for the radius of a $1.4M_{\odot}$ star (see Fig.~\ref{MRplot}), $R(1.4 M_\odot) > 9.9\; {\rm km}$.
We further note that this bound is unaffected by the conclusions of \cite{Bauswein:2017vtn}, which constrain the minimum radius of a NS given the existence of the kilonova associated with GW170817.

 %%%%%%%%%%%%%%%%%%%%%%%%%%%%%%%%%%%%%%%%%%%%%%%%%%%%%%%%%%%%
\begin{figure}
\includegraphics[width = 0.48\textwidth]{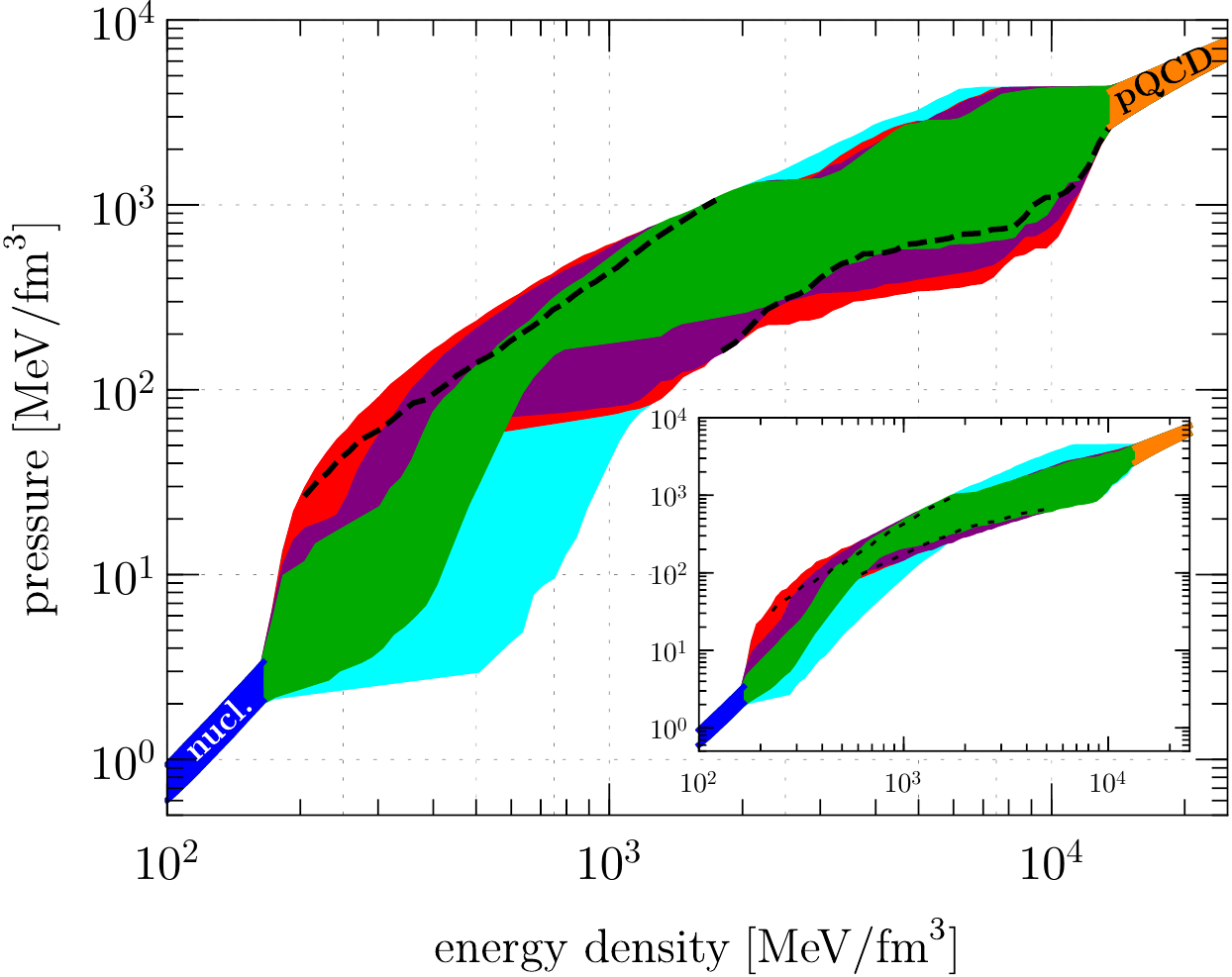}
\caption{Our ensemble of EoSs shown in the form of $\epsilon$ vs.~$p$. The color coding follows that of the previous figures, with the addition of a blue region indicating the nuclear EoSs and an orange region indicating the pQCD EoS. The black dashed lines indicate where the upper and lower edges become truncated with a further restriction of $M_{\text{max}} < 2.16 M_{\odot}$ (see \cite{Margalit:2017dij,Rezzolla:2017aly,Ruiz:2017due}). Inset: the same function constructed with tritropic interpolating functions only. \label{epsvsp}}
\label{index}
\end{figure}
%%%%%%%%%%%%%%%%%%%%%%%%%%%%%%%%%%%%%%%%%%%%%%%%%%%%%%%%%%%%
 
\subsection{Constraints on the Equation of State}

In addition to the macroscopic observables discussed above, we also study the effects of the astrophysical constraints on the EoS itself. This is done in Fig.~\ref{epsvsp}, where we display our family of EoSs in the energy density vs.~pressure plane. Here, we see how the $2 M_{\odot}$ constraint excludes EoSs that are soft at low densities, while the $\Lambda$ constraint excludes EoSs that are more stiff at low densities. This is of course natural, considering that the latter EoSs are the ones that produce stars with large radii and thereby also large $\Lambda$. Also in this figure, we display black dashed lines to indicate where the upper and lower edges become truncated with a further restriction of $M_{\text{max}} < 2.16 M_{\odot}$, a bound proposed in \cite{Margalit:2017dij,Rezzolla:2017aly,Ruiz:2017due} stemming from the resulting kilonova observations associated with GW170817. 

The EoS bounds can be quantified by inspecting the effects of the astrophysical observations on the EoS parameters. The parameter that is physically the most meaningful is clearly $\gamma_{1}$, whose allowed values we show for the tritropic and quadrutropic EoSs in Table I. Restricting ourselves here to those EoSs where the first polytropic interval extends to a density $n \geq 1.5 n_{s}$, so that it is of non-negligible size, the range of $\gamma_1$ becomes $0.05 < \gamma_{1} < 8.5$.  Imposing the $2 M_{\odot}$ condition further leads to the lower limit increasing to $\gamma_{1} > 0.6$, while the constraint $\Lambda(1.4 M_{\odot}) < 800$ reduces the upper limit to $\gamma_{1} < 6.7$. It is interesting to note that this combined limit of $0.6 < \gamma_{1} < 6.7$ is in rough agreement with the (in principle ad hoc) requirement enforced upon the same quantity in \cite{Hebeler:2013nza}, $1 < \gamma_{1} < 4.5$. We emphasize, however, that our bound is based solely on quantifiable theoretical and observational constraints.

\subsection{Robustness of the results}

To gauge the robustness of the results described above, there are two issues to consider. The first concerns the sensitivity of our findings to the number of interpolating polytropes, which can be estimated by comparing results obtained with three and four polytropes, respectively. This is indeed done Fig.~\ref{epsvsp}, where the inset of the figure shows the EoS family that results from tritropic interpolation. We observe that upon imposing the two-solar-mass constraint, the two results are in good quantitative agreement. The most significant difference can be witnessed at low densities, where the fourth polytrope allows a small number of EoSs that are initially softer or stiffer than what would be feasible with tritropic interpolation. On the mass-radius plane, these findings most importantly translate to the appearance of stars with relatively small radii, $R(1.4 M_\odot)\lesssim 10$ km. In addition, we observe that adding the fourth polytrope allows for some light stars with $M < 1.4 M_{\odot}$ that have larger radii than what is allowed by the tritropic interpolation. These configurations correspond to EoSs that are initially stiff but undergo a rapid qualitative change and become soft already at rather low densities $n < 1.5 n_s$. This region is excluded if we assume that the first polytrope continues to a density $n \geq 1.5n_s$, as done in \cite{Hebeler:2013nza}.

Although we cannot make a firm statement without a direct computation, we suspect that including a fifth polytropic segment would not change our conclusions appreciably. In \cite{Raithel:2016bux}, it was found that a polytropic function consisting of five segments of even spacing suffices to reproduce all realistic EoSs. In our case, the lengths of the polytropic segments are varied, so that even with the nuclear and pQCD constraints, our quadrutropic interpolation function has the same number of free parameters as the ansatz used in this reference.

%%%%%%%%%%%%%%%%%%%%%%%%%%%%%%%%%%%%%%%%%%%%%%%%%%%%%%%%%%%%
\begin{table}
\begin{tabularx}{0.48\textwidth}{cYYYY} \toprule
3-tropes& All EoSs & $2 M_\odot$ & $\Lambda< 800 $ & $\Lambda < 400$ \\
\midrule
$\gamma_1$	& 0.2-8.5 & 0.7-8.5 & 0.7-6.6 & 0.7-4.7 \\
$M_{\rm max} [M_\odot]$  & $<$0.5-3.0 & 2.0-3.0 & 2.0-2.7 & 2.0-2.3 \\
$R(1.4M_\odot) [{\rm km}]$ 	& 7.1-14.6 & 10.7-14.6 & 10.7-13.6 & 10.7-12.4 \\
\midrule
4-tropes& All EoSs & $2 M_\odot$ & $\Lambda< 800 $ & $\Lambda < 400$ \\
\midrule
$\gamma_1$	& 0.05-8.5 & 0.6-8.5 & 0.6-6.7 & 0.6-4.7 \\
$M_{\rm max} [M_\odot]$  & $<$0.5-3.2& 2.0-3.2 & 2.0-3.0 & 2.0-2.5 \\
$R(1.4M_\odot) [{\rm km}]$ 	& 6.6-14.6 & 9.9-14.6 & 9.9-13.6 & 9.9-12.5 \\
\bottomrule
\end{tabularx}
\caption{Allowed parameter values for our tritropic and quadrutropic solutions, arising from the matching procedure. The first column corresponds to all thermodynamically-consistent EoSs, the second to those fulfilling the 2$M_{\odot}$ constraint, and the last two to those that additionally satisfy $\Lambda(1.4M_\odot)< 800$ and $\Lambda(1.4M_\odot)< 400$, respectively. For the $\gamma_{1}$ row only, we impose the extra requirement that the first polytropic segment last until at least $n=1.5n_s$, so that $\gamma_{1}$ may carry robust physical meaning.}
\end{table}
%%%%%%%%%%%%%%%%%%%%%%%%%%%%%%%%%%%%%%%%%%%%%%%%%%%%%%%%%%%%

\begin{figure}[t]
\includegraphics[width=0.48\textwidth]{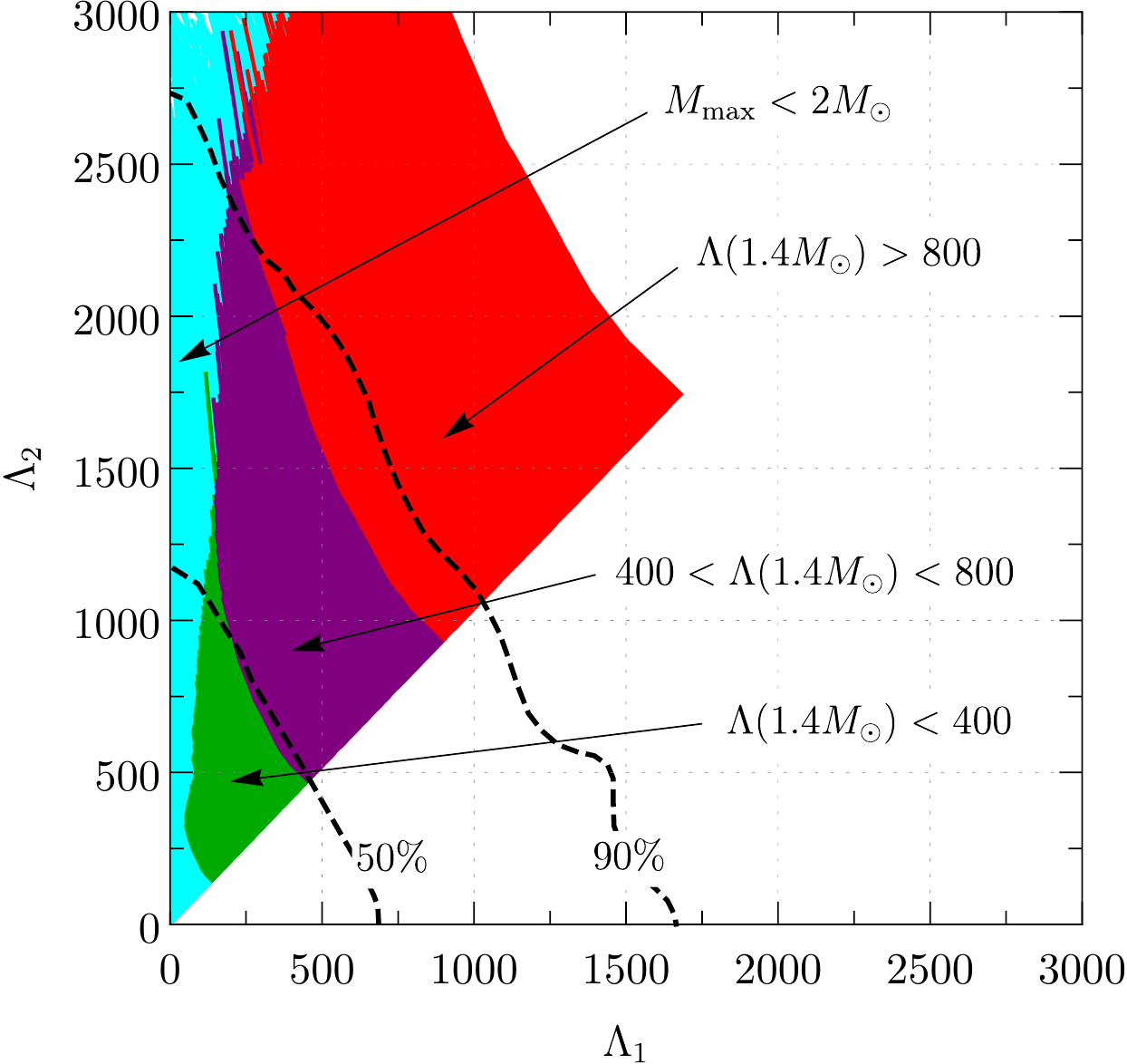}
\caption{A comparison of our three sets of quadrutropic EoSs, corresponding to different values of $\Lambda(1.4 M_\odot)$, with the 90\% and 50\% probability contours given in Fig.~5 of \cite{TheLIGOScientific:2017qsa}. The color coding follows that of the previous figures.}
\label{L-L}
\end{figure}

Another question to inspect is whether our choice of enforcing the tidal-deformability constraint as a limit for a 1.4$M_\odot$ star leads to results different from those we would have obtained using the other forms of data provided in \cite{TheLIGOScientific:2017qsa}. In particular, in Fig.~\ref{L-L} we reproduce the 90\% and 50\% probability contours for the tidal deformabilities of the two stars measured by LIGO and Virgo, given in Fig.~5 of \cite{TheLIGOScientific:2017qsa}. Alongside these contours, we show regions composed of our EoSs, which are generated by varying the mass of one of the two stars within the uncertainty region reported in \cite{TheLIGOScientific:2017qsa} and solving for the other using the accurately-known chirp mass of the merger, $\mathcal{M}=1.188 M_{\odot}$. Inspecting the boundaries of the colored regions of this figure, corresponding to different $\Lambda(1.4 M_\odot)$ values in the same fashion as in our earlier figures, we observe good qualitative agreement with the 90\% and 50\% probability contours of LIGO and Virgo. More quantitatively, if we were to use the 90\% probability contour as an exclusion bound in place of the condition $\Lambda(1.4 M_{\odot}) < 800$, we would arrive at the constraint $R(1.4M_\odot) \in [9.9,13.8]$ km. This demonstrates the robustness of our conclusions with respect to the way the $\Lambda$ limit is implemented.

%%%%%%%%%%%%%%%%%%%%%%%%%%%%%%%%%%%%%%%%%%%%%%%%%%%%%%%%%%%%
%%%%%%%%%%%%%%%%%%%%%%%%%%%%%%%%%%%%%%%%%%%%%%%%%%%%%%%%%%%%

%%%%%%%%%%%%%%%%%%%%%%%%%%%%%%%%%%%%%%%%%%%
\section{Conclusions}

The simultaneous observation of gravitational and electromagnetic signals from the merger of two compact stars has recently begun a new era of multimessenger astronomy for NSs \cite{TheLIGOScientific:2017qsa,Coulter:2017wya,GBM:2017lvd,Monitor:2017mdv,Cowperthwaite:2017dyu,Soares-Santos:2017lru,Nicholl:2017ahq,Margutti:2017cjl,Chornock:2017sdf,Radice:2017lry} and has opened up a completely new window on the properties of the strongly-interacting matter inside them (see also \cite{Margalit:2017dij,Rezzolla:2017aly,Ruiz:2017due,Radice:2017lry,Bauswein:2017vtn,Gupta:2017vsl,Zhou:2017xhf,Posfay:2017cor,Lai:2017mjv}). In particular, the upper limit placed on $\Lambda$ by LIGO and Virgo constrains the stiffness of the matter within these objects. This has far-reaching implications for both the EoS of nuclear matter and the macroscopic properties of NSs, which we have quantitatively studied in the present paper. 

The main conclusion of our work is that we are entering an age where astrophysical measurements are beginning to set extremely stringent bounds on the collective properties of dense QCD matter. This can be easily witnessed, even with the naked eye, from our Fig.~\ref{epsvsp}: the tidal-deformability measurement alone is enough to significantly decrease the uncertainty in the NS-matter EoS. While our discussion has mostly concentrated on astrophysical bounds as well as the EoS at relatively low densities, in the future it will be interesting to ask whether astrophysical constraints can even lead to robust statements about the existence of quark matter inside NS cores or whether quantitative bounds can be set on the properties of high-density quark matter. These are amongst the questions that will be studied by us in future works, also including input from direct radius measurements \cite{Bauswein:2017vtn,Radice:2017lry,Yamamoto:2017wre,Guillot:2013wu,Bogdanov:2016nle,Steiner:2017vmg,Nattila:2015jra,Nattila:2017wtj,Watts:2016uzu,Ozel:2016oaf}.
\acknowledgments{
We thank Kai Hebeler, Joonas N\"attil\"a, Juri Poutanen, and Achim Schwenk for useful discussions. The work of EA, TG, and AV has been supported by the Academy of Finland, Grant no.~1303622 as well as by the European Research Council, grant no.~725369. In addition, EA gratefully acknowledges support from the Finnish Cultural Foundation.
}

%%%%%%%%%%%%%%%%%%%%%%%%%%%%%%%%%%%%%%%%%%%%%%%%%%%%%%

\end{document}